\documentclass[secnumarabic,amssymb, nobibnotes, aps, prc,twocolumn,showkeys]{revtex4} 

\usepackage{graphicx}
\usepackage{amssymb}
\usepackage{amsmath}
\usepackage{bm}
\usepackage{kotex}
\usepackage{times}
\usepackage{xcolor}
\usepackage{ulem}
\usepackage{cancel}
\usepackage{verbatim}
\usepackage{pifont}
\usepackage{rotating}
\usepackage{subfigure}
\usepackage{multirow}
\usepackage{changepage}
\usepackage{mathtools}

\usepackage[colorlinks=true,linkcolor=red,citecolor=blue]{hyperref}

\newcommand{\be}{\begin{equation}}
\newcommand{\ee}{\end{equation}}
\newcommand{\bea}{\begin{eqnarray}}
\newcommand{\eea}{\end{eqnarray}}

\begin{document}
\setcounter{page}{1}

\title{Correlation between alpha-decay half-lives and symmetry energy}

\author{Yong-Beom Choi}
    \email{1991.yb.choi@gmail.com}
    \affiliation{Center for Innovative Physicist Education and Research, Extreme Physics Institute, and Department of Physics, Pusan National University, Busan 46241, Korea}
    \affiliation{Department of Physics Education, Daegu University, Gyeongsan 38453, Korea}
\author{Hana Gil} 
    \affiliation{Center for Extreme Nuclear Matters, Korea University, Seoul 02841, Korea}
\author{Chang Ho Hyun}
    \affiliation{Department of Physics Education, Daegu University, Gyeongsan 38453, Korea}
\author{Chang-Hwan Lee}
    \email{clee@pusan.ac.kr}
    \affiliation{Center for Innovative Physicist Education and Research, Extreme Physics Institute, and Department of Physics, Pusan National University, Busan 46241, Korea}

\date[]{}

\begin{abstract}
We study the alpha-decay half-lives of $84 \le Z  \le 92$ even-even nuclei in the semiclassical 
WKB approximation framework using the density-dependent cluster model and the density
distribution described by various Korea-IBS-Daegu-SKKU (KIDS) models. Main goal
of this work is to find a correlation between the alpha-decay half-lives and the stiffness of
the symmtery energy. Parameters of KIDS model are determined to reproduce the
nuclear data (energy and charge radii of $^{40}$Ca, $^{48}$Ca and $^{208}$Pb) and the neutron star
observations including constraints by NICER, tidal deformability from GW170817, and
the maximum mass limit of neutron star observations. We use the KIDS models (A-D)
which have distinctive values for the parameters controlling the stiffness of the nuclear
matter equation of state. We find that the alpha-decay half-lives increase systematically
from KIDS-A model to KIDS-D model. This implies a correlation between the alpha-decay
half-lives and the symmetry energy. 
We find that the correlation can be understood from the dependence of the particle distribution in the core and the surface region on the symmetry energy.

\end{abstract}


\maketitle



\section{Introduction}

The detection of gravitational waves GW170817 and the observation of electromagnetic afterglows which were generated by the merger of a neutron star binary opened a new era of multi-messenger astrophysics~\cite{LIGOScientific:2017vwq}.
Tidal deformability from GW170817 allows us to constrain neutron star equations of state using gravitational waves~\cite{LIGOScientific:2017ync}. 
In addition, simultaneous measurement of masses and radii of neutron stars  by NICER
provides challenging constraints on the neutron star equation of state~\cite{Riley:2019yda, Miller:2019cac, Riley:2021pdl, Raaijmakers:2021uju}. 
One of the key ingredients of neutron star equations of state is the symmetry energy of dense nuclear matter which characterizes the change of the binding energy as the neutron to proton ratio of a nuclear system changes. 
Symmetry energy plays a critical role for the physical quantities in
nuclear physics and astrophysics, such as structures of nuclei, heavy-ion reactions, and properties of neutron stars~\cite{Baldo:2016jhp, Lattimer:2014scr}.

Alpha ($\alpha$) decay remains an important process in nuclear physics, providing rich information on nuclear structure and dynamics, and improving the predictability of contemporary nuclear models used in nuclear astrophysics calculations~\cite{Buck:1992zz, Li:2009sr, Deng:2017ids, Sun:2016bbw}.
Alpha decay is also important for the experimental identification of new nuclides in heavy and superheavy regions~\cite{Nurmukhamedov:2018yqu, Qian:2014pbn}.
Various theoretical models have been developed to explain alpha decay within the framework of quantum tunneling and to provide a good description of experimental alpha-decay half-lives~\cite{Choi:2023ztx}. The relationship between the alpha-decay half-lives and the symmetry energy in nuclear physics is a complex problem.
Different parameterizations of the Skyrme energy density functionals have been used to characterize the symmetry energy coefficient and its density-slope parameter, providing insights into the relationship between the symmetry energy and the alpha-decay half-lives~\cite{Seif:2018nqn}.

In this work, we study the alpha-decay half-lives of $84 \le Z  \le 92$ even-even nuclei in the semiclassical WKB approximation framework using the density-dependent cluster model and the density distribution described by various Korea-IBS-Daegu-SKKU (KIDS) models~\cite{Gil:2020wct, Gil:2023swx, Gil:2021cmp}. In KIDS model, energy density of many-nucleon system is expanded in the power of the Fermi momentum. Since the coefficients in KIDS model cannot be calculated from the first principles, they must be determined empirically and the optimal number of coefficients is chosen to describe the infinite nuclear matter. 
We use the KIDS models (A-D) which have distinctive values for the parameters controlling the stiffness of the nuclear matter equation of state~\cite{Gil:2020wct}.
Parameters of KIDS model are determined to reproduce the
energy and charge radii of $^{40}$Ca, $^{48}$Ca and $^{208}$Pb, and the neutron star
observations including constraints by NICER, tidal deformability from GW170817, and
the maximum mass limit of neutron star observations ~\cite{Gil:2020wct,Hinderer:2009ca, LIGOScientific:2018cki}.  

Main goal of this work is to find a correlation between the alpha-decay half-lives and the stiffness of
the symmetry energy.
We found that the alpha-decay half-lives increase systematically
from KIDS-A model to KIDS-D model. This implies a correlation between the alpha-decay
half-lives and the symmetry energy.

The paper is organized as follows.
In Sec.~\ref{sec_alpha_decay}, the WKB approximation for calculating the alpha-decay half-lives is described. 
We use the cluster-formation model~\cite{SalehAhmed:2013nwa, Ahmed:2015kra, Deng:2015qha, Deng:2016ibo} to estimate the preformation factor $P_\alpha$, the probability of the preformed $\alpha$ particle inside the parent nucleus. 
The density distributions of axially deformed nuclei are treated by expanding the densities in terms of the Legendre polynomials.
The alpha-decay half-lives of axially deformed nuclei are calculated under the density-dependent cluster model~\cite{Xu:2005ukj, Xu:2006fq, Ni:2008zza} by considering the density distributions.
In Sec.~\ref{sec_KIDSmodel}, we introduce the KIDS model and the selected parameters to reproduce the nuclear data and the neutron star observations. 
The selected KIDS models show the distinct stiffness of the nuclear matter equation of state.
For chosen KIDS models, distributions of proton and neutron are discussed. 
In Sec.~\ref{sec_result}, to confirm the validity of the selected KIDS models, we first compare the binding energies and quadrupole deformation described by KIDS models with the experimental data. 
We then calculate the alpha-decay half-lives using selected KIDS models.
To understand the origin of the correlation between the symmetry energy and the alpha-decay half-lives, we compare the calculated alpha-decay half-lives and the potential between daughter nucleus and $\alpha$ particle from the KIDS models.
In Sec.~\ref{summary_and_conclusions}, we summarize our results and discuss their implications.


\section{WKB approximation for alpha-decay half-lives}
\label{sec_alpha_decay}

Using the WKB approximation, the alpha-decay half-life and the decay width are calculated by using~\cite{Xu:2005ukj, Gurvitz:1986uv}
\begin{eqnarray}
T_{1/2} &=& \frac{\hbar \ln 2}{\Gamma}, \\
\Gamma &=& P_\alpha N_f \frac{\hbar^2}{4\mu} P_{\rm tot}. \label{eq:Gamma}
\end{eqnarray}
In cluster-formation model~\cite{SalehAhmed:2013nwa, Ahmed:2015kra}, a preformation factor $P_\alpha$ is defined as
\begin{equation}
P_\alpha = \frac{2 S_p + 2 S_n - S_\alpha}{S_\alpha},
\end{equation}
where $S_p$, $S_n$ and $S_\alpha$ denote the separation energies for the one proton, one neutron and the one alpha particle,
respectively.
$\mu$ is the reduced mass of the alpha particle and daughter nucleus, and the normalization factor $N_f$ and total penetration probability $P_{\rm tot}$
are given as
\begin{eqnarray}
N_f &=& \frac{1}{2} \int^\pi_0 N_f (\beta) \sin \beta d\beta, \\
P_{\rm tot} &=& \frac{1}{2} \int^\pi_0 \exp \left[ -2 \int^{r_3(\beta)}_{r_2(\beta)} k(r', \beta) dr' \right] \sin \beta d\beta,
\end{eqnarray}
with 
\begin{eqnarray}
N_f(\beta) &\simeq& \left[ \int^{r_2(\beta)}_{r_1(\beta)} \frac{dr'}{2 k(r', \beta)} \right]^{-1}, \label{eq:nfapp} \\
k(r, \beta) &=& \sqrt{\frac{2\mu}{\hbar^2} |Q_\alpha - V(r, \beta)|}.
\end{eqnarray}
$r_1$, $r_2$ and $r_3$ are the distances for the classical turning points in the order $r_1 < r_2 < r_3$.
In density-dependent model~\cite{Xu:2005ukj}, potential $V(r,\beta)$ consists of the centrifugal with the Langer modified form~\cite{Langer:1937qr, Buck:1992zza, Buck:1993sku}, Coulomb and nuclear potentials
\begin{eqnarray}
V(r, \beta) &=& V_l (r) + V_C(r, \beta) + V_N(r, \beta), \\
V_l(r) &=& \frac{\hbar^2}{2\mu} \frac{(l+1/2)^2}{r^2}, \\
V_C(r, \beta) &=& \int d\bm{r}_d d\bm{r}_\alpha \rho^p_d(\bm{r}_d) \rho^p_\alpha (\bm{r}_\alpha) \frac{e^2}{s},
\label{coulV}\\
V_N(r, \beta) &=& \lambda \int d\bm{r}_d d\bm{r}_\alpha \rho_d(\bm{r}_d) \rho_\alpha (\bm{r}_\alpha) v(\bm{s}). 
\label{nuclV}
\end{eqnarray}
The quantum number $l$ is the orbital angular momentum carried by alpha particle. In this study, $l=0$ is applied since only even-even nuclei are considered \cite{Xu:2006fq}.
$\rho_{\alpha,{\rm d}}$ and $\rho_{\alpha, {\rm d}}^p$ represent the nucleon and proton distribution within $\alpha$ and daughter (${\rm d}$) nucleus, respectively, 
and the relative coordinate $\bm{s}$ is defined as $\bm{s} = \bm{r} + \bm{r}_\alpha - \bm{r}_{\rm d}$.
For the alpha particle, we adopt the standard Gaussian form $\rho_\alpha(r) = 0.4229 \exp(-0.7024 r^2)$~\cite{Satchler:1979ni}.
$v(r)$ is the effective nucleon-nucleon interaction, for which we use the M3Y model given by~\cite{Satchler:1979ni,  Bertsch:1977sg,  Kobos:1982pdw}
\begin{eqnarray}
v(\bm{s}) = 7999 \frac{e^{-4s}}{4s} - 2134 \frac{e^{-2.5 s}}{2.5 s} - 276 \left( 1 - 0.005 \frac{Q_\alpha}{A_\alpha} \right)\delta(\bm{s}), \nonumber\\
\end{eqnarray}
where the units of $v(\bm{s})$ and $Q_\alpha$ are MeV and $A_\alpha$ is the mass number of alpha particle.

Deformations of the nucleus are accounted by expanding Coulomb and nuclear potentials $V_I$ ($I=C, N)$ in terms of the Legendre polynomial 
$P_n(\cos \beta)$ as~\cite{1983ZPhyA.310..287R, Bai:2018hbe}
\begin{eqnarray}
V_I(r, \beta) &=& \sum_{n=0, 2, 4, \cdots} V_{I n}(r,\beta), \\
V_{Cn}(r, \beta) &=& 8 e^2 \int^\infty_0 \!\! dk j_n(k r) \tilde{\rho}^p_\alpha(k) \nonumber \\
&\times& \left[\int^\infty_0\!\!  dr' r'^2  \rho^p_{{\rm d} n}(r') j_n(kr') \right] \! P_n(\cos \beta), \\
V_{Nn}(r, \beta) &=& \frac{2}{\pi} \int^\infty_0 \!\! dk k^2 j_n(k r) \tilde{\rho}_\alpha(k) \tilde{v}(k) \nonumber \\
&\times& \left[\int^\infty_0\!\! dr' r'^2  \rho_{{\rm d} n}(r') j_n(kr') \right] \! P_n(\cos \beta),
\end{eqnarray}
where the tilded quantities denote the Fourier transformation of the corresponding quantity in coordinate space.

Normalization factor $\lambda$ for the nuclear potential in Eq.~(\ref{nuclV}) is determined by the Bohr-Sommerfeld quantization condition~\cite{Xu:2005ukj}
\begin{eqnarray}
\int^\pi_0\!\! \int^{r_2(\beta)}_{r_1(\beta)}\!\!\! \sqrt{\frac{2\mu}{\hbar^2} |Q_\alpha - V(r, \beta)|} \sin\beta dr d\beta
= (2n+1)\frac{\pi}{2} \nonumber \\
= (G-l+1)\frac{\pi}{2}.
\end{eqnarray}
In the initial calculation, global quantum number $G$ was assumed to be $G=20$ for $82 < N_{\rm par} \leq 126$ 
and $G=22$ for $N_{\rm par} >126$ as in Ref.~\cite{Choi:2023ztx}.
However, since the obtained results overshoot the experimental data,
we adjusted the $G$ values to have a minimal deviation from experimental data.
The result is being discussed after the models are introduced.

\section{KIDS Model}
\label{sec_KIDSmodel}

Main goal of the work is to find out whether there exists a correlation between the alpha-decay half-lives and the stiffness of the symmetry energy, and if it does, how they are related. 
In order to satisfy the purpose, models that have different density dependence of the symmetry energy are necessary.
KIDS-A, B, C, and D models in Ref. \cite{Gil:2020wct} are determined to reproduce the nuclear data and the neutron star observations.
More specifically, energy and charge radius of $^{40}$Ca, $^{48}$Ca and $^{208}$Pb are used for the nuclear data,
and radius of $1.4 M_\odot$-mass neutron star contrained by NICER, tidal deformability determined from GW170817,
and maximum mass limit $M_{\rm max} \geq 2 M_\odot$ are used to constrain the incompressibility of the symmetric matter $K_0$
and the symmetry energy parameters $J$, $L$ and $K_{\rm sym}$ in the expansion of the symmetry energy
\[
S(\rho) = J + L x + \frac{1}{2}K_{\rm sym} x^2 + \cdots,
\]
where $x = (\rho - \rho_0)/3\rho_0$ with $\rho_0$ the saturation density.
Numerical values of $K_0$, $J$, $L$ and $K_{\rm sym}$ for the KIDS-A, B, C and D models are summarized in Tab. \ref{tab1}.
Models have distinctive values for the parameters that control the stiffness of the nuclear matter equation of state.
The models have been used to calculate the properties of even-even nuclei from the proton drip line to the neutron drip line,
and it is shown that they are good at reproducing the well-known measured properties across the nuclear chart,
and the accuracy of the four models is similar to that of the SLy4 model \cite{Gil:2023swx}.
At the same time, the models show evident model dependence in the neutron rich region, i.e. close to the neutron drip line.
Nuclei we consider in the work are Po, Rn, Ra, Th, U, for which $N-Z = 22$--54 or the relative asymmetry $(N-Z)/A = 0.115$--0.243,
so the neutron number overwhelms the proton number in all the nuclei.
Therefore, KIDS-A, B, C, and D models offer an adequate platform to examine systematically the effect of the symmetry energy
to the alpha-decay half-lives of heavy isotopes.

\begin{table}
\centering
\begin{tabular}{ccccc}\hline
 Model &\phantom{a} $K_0$ \phantom{a}&\phantom{a} $J$ \phantom{a}&\phantom{a} $L$ \phantom{a}
& \phantom{a} $K_{\rm sym}$ \phantom{a}\\ \hline
KIDS-A & 230 & 33 & 66 & $-139.5$ \\
KIDS-B & 240 & 32 & 58 & $-162.1$ \\
KIDS-C & 250 & 31 & 58 & $-91.5$ \\
KIDS-D & 260 & 30 & 47 & $-134.5$ \\ \hline
\end{tabular}
\caption{The selected parameter values for the KIDS models. Units are in MeV.}
\label{tab1}
\end{table}

\begin{figure}
	\centering
	\includegraphics[width=0.45\textwidth]{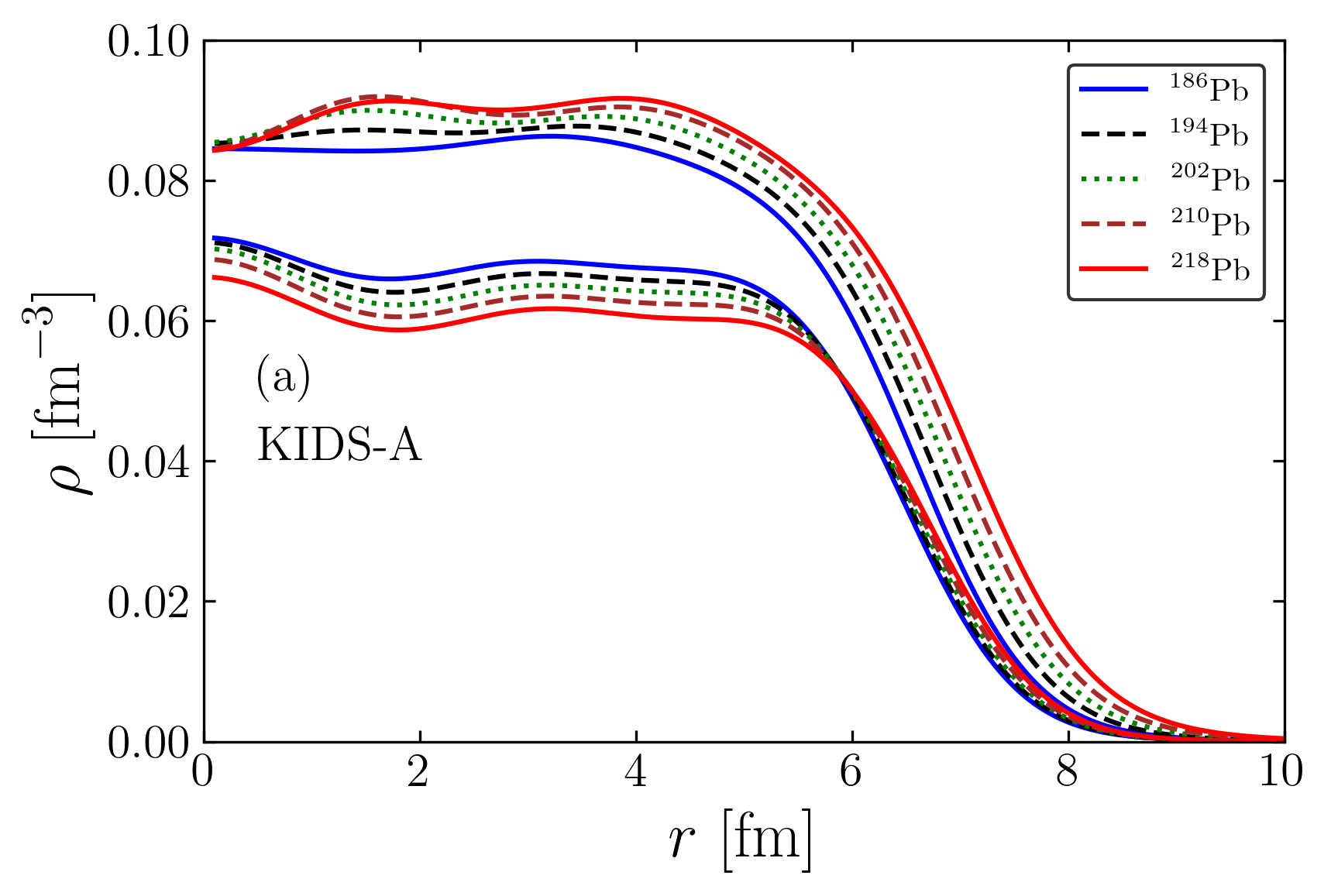}
	\includegraphics[width=0.45\textwidth]{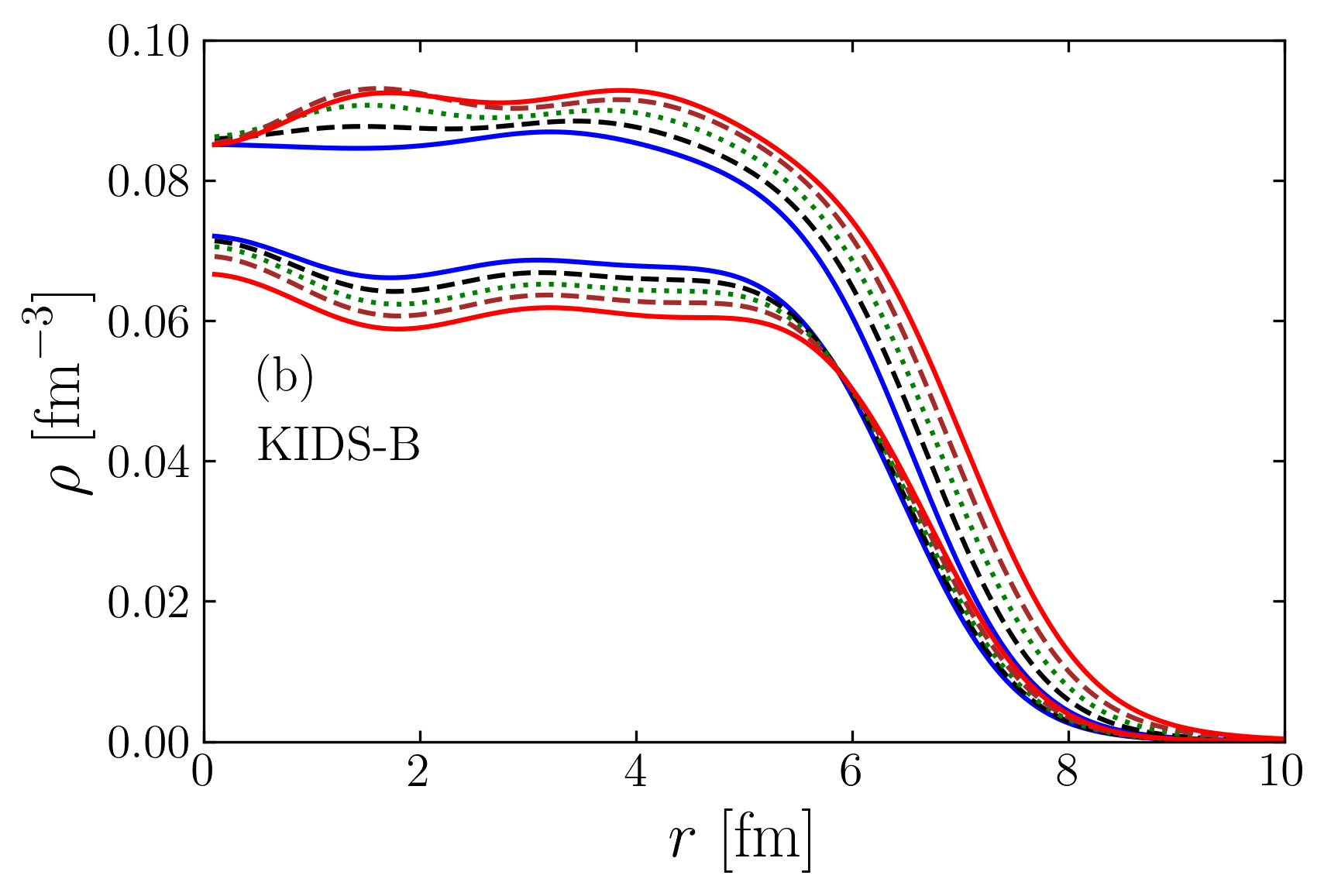}
	\includegraphics[width=0.45\textwidth]{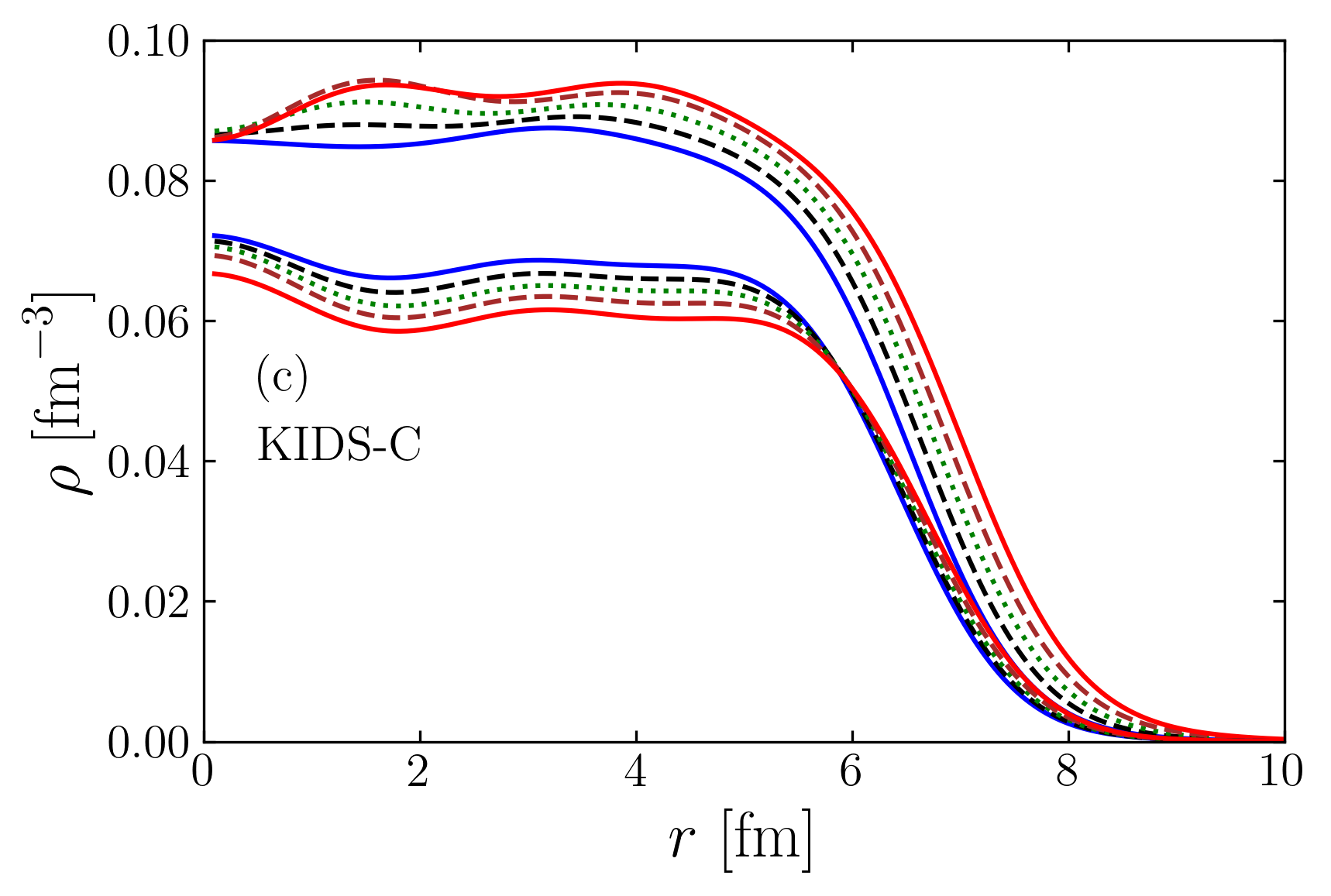}
	\includegraphics[width=0.45\textwidth]{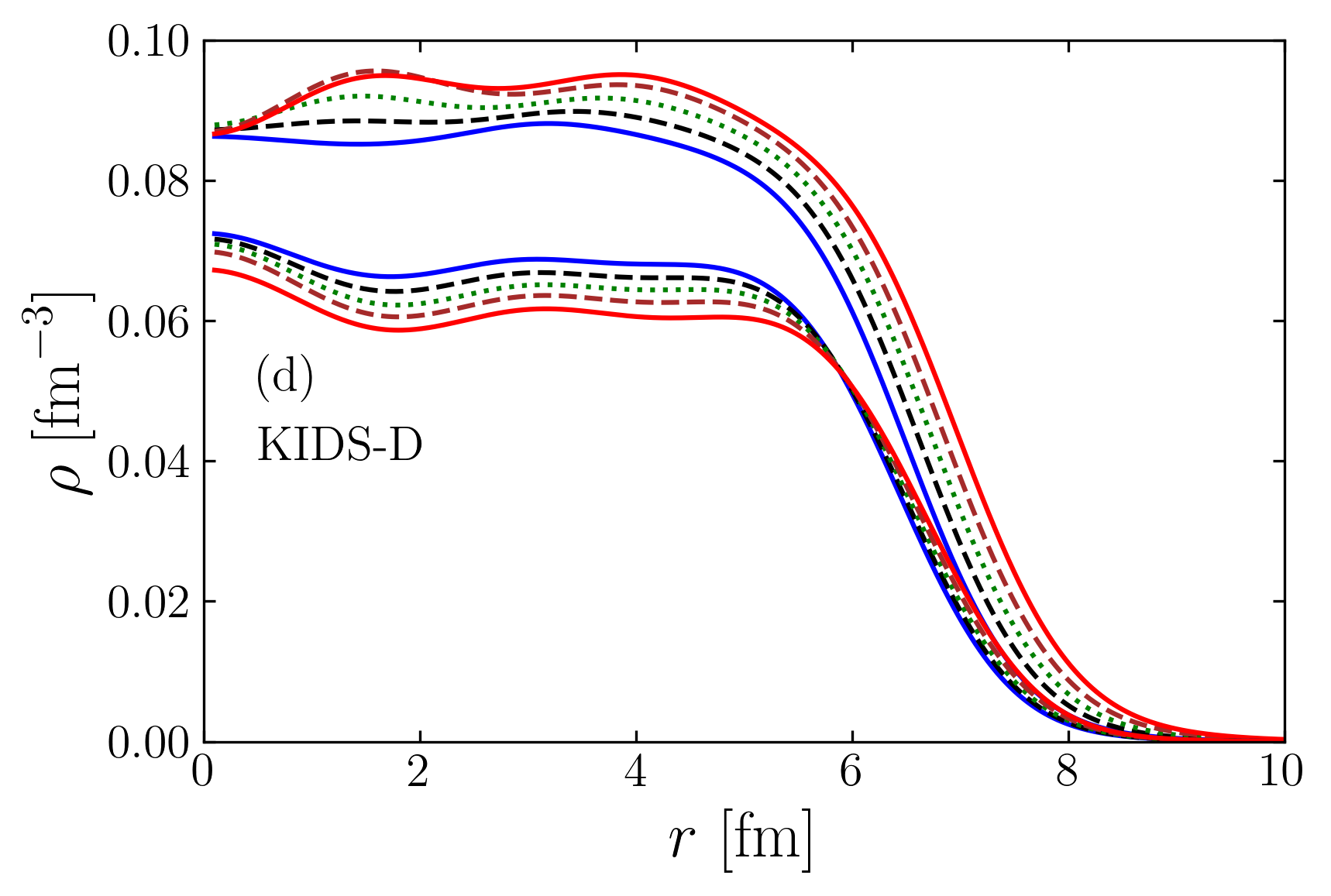}
	\caption{Proton (curves in the lower group) and neutron (curves in the upper group) distributions in the Pb isotopes with KIDS-A, B, C and D models.}
	\label{fig1}
\end{figure}

Figure \ref{fig1} shows the proton and the neutron distributions in the Pb isotopes with the KIDS-A, B, C and D models.
In the core region where the distribution is flat as a whole, proton density is suppressed as the mass number $A$ increases,
and it is opposite for the neutron distribution.
Behavior of the neutron is easily understood because the neutron number increases as $A$ increases.
Depletion of the proton in the core for a larger $A$ is compensated by the enhancement in the surface region $r=6$--8 fm.
Difference in the proton distribution for different isotopes can have a direct effect to the Coulomb potential given by Eq. (\ref{coulV}).

Considering the model dependence, the proton distribution is almost identical within the models, which
confirms that the proton distribution is insensitive to the symmetry energy.
However, one can see that the neutron distribution is strongly dependent on the symmetry energy by looking at the distribution in the core.
There is a maximum slightly below $r=2$ fm, and the width of the band at the maximum is wider for KIDS-D than KIDS-A.
It means that as the symmetry energy is soft, neutron distribution in the core is enhanced, which also affects the neutron distribution
in the surface and the tail regions.
Neutron distribution contributes to the nuclear potential in Eq. (\ref{nuclV}) by means of $\rho_{\rm d}(\bm{r}_{\rm d})$, so the dependence on the symmetry energy can have effects to the location, height and width of the tunneling barrier.
As will be seen in the next section, the dependence on the symmetry energy has a systematic effect to the alpha-decay half-lives.

\section{Result}
\label{sec_result}

In this work we consider five parent nuclides :
Po ($Z=84$) isotopes in the mass range $A=190$--224,
Rn ($Z=86$) isotopes in the mass range $A=194$--230,
Ra ($Z=88$) isotopes in the mass range $A=202$--234,
Th ($Z=90$) isotopes in the mass range $A=210$--238, and
U ($Z=92$) isotopes in the mass range $A=216$--242.
We take the $\alpha$-decay energy $Q_\alpha$ from AME 2020~\cite{Wang:2021xhn}.
Binding energies of odd-$A$ nuclei are not available yet in the KIDS model. 
Therefore, we also take the binding energies of the nuclei from AME2020 to estimate $P_\alpha$.

\begin{figure}
	\centering
    \includegraphics[width=0.45\textwidth]{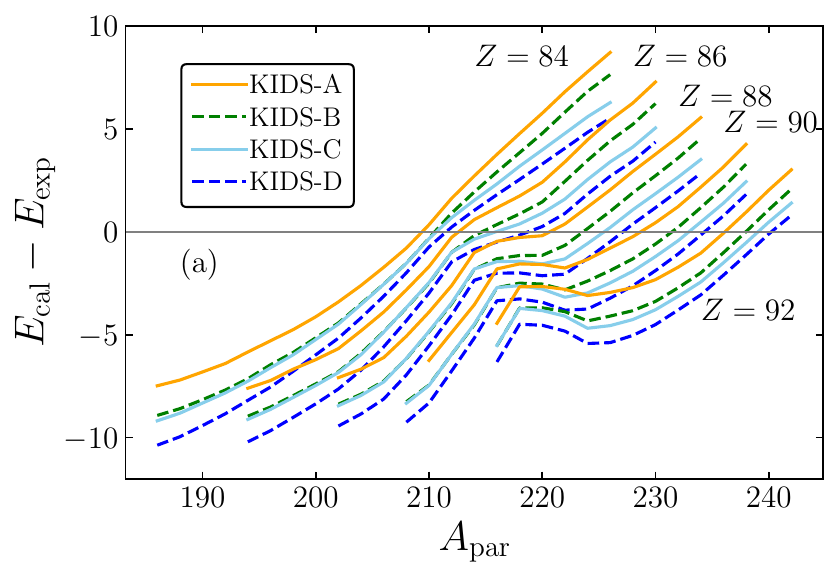}
    \includegraphics[width=0.45\textwidth]{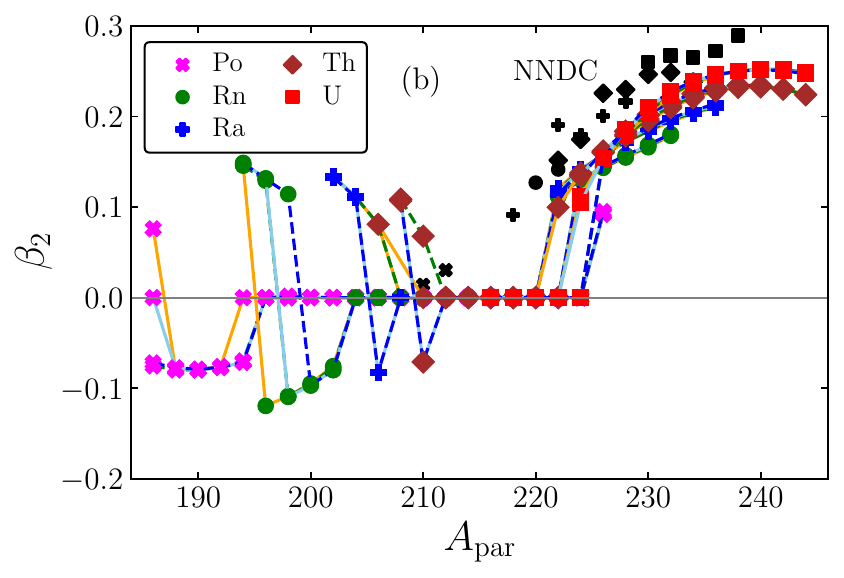}
	\caption{(a) Difference of the binding energy between theory and experiment. (b) Quadrupole deformation.}
	\label{fig2}
\end{figure}

To verify the accuracy of the model, we calculate the difference of the binding energy between theory and experiment~\cite{Wang:2021xhn} $E_{\rm cal} - E_{\rm exp}$,
and compare the quadrupole deformation $\beta_2$ with the data from NNDC~\cite{NNDC}.
Fig. \ref{fig2}(a) shows the binding energy difference for the isotopes considered in the work.
For all the isotopes and for all the models, the difference tends to increase as the mass number of the parent nuclei $A_{\rm par}$ increases.
For a given $Z$ number, the line of KIDS-A is located in the highest place,
and it becomes lower in the order of KIDS-B, KIDS-C and KIDS-D.
These systematic changes of the energy difference are consistent with the results of the Nd isotopes with the KIDS-A, B, C, and D models \cite{Gil:2021cmp}.
It is also notable that the range of difference becomes narrow as 
the $Z$ value increases.

Fig. \ref{fig2}(b) displays the quadrupole deformation $\beta_2$.
Models are distinguished by colors of the line and nuclides by symbols and give results notably different from each other in the interval $194 \leq A_{\rm par} \leq 210$.
At the other $A_{\rm par}$ values, model predictions are consistent and reproduce the data fairly well.
By and large, both energy and deformation agree with experimental data with reasonable error ranges,
so the models are ready to be applied to the alpha-decay.

\begin{figure}
	\centering
    \includegraphics[width=0.45\textwidth]{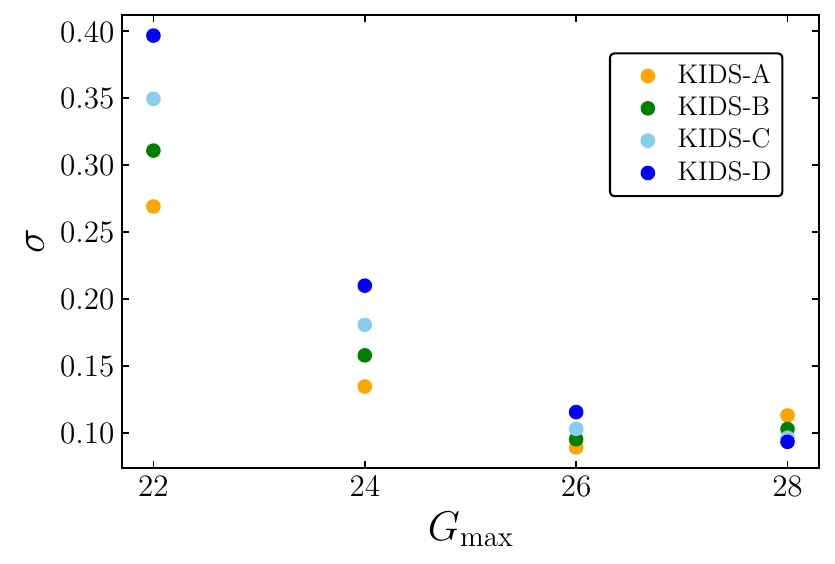}
	\caption{Root-mean-square deviation $\sigma$ of the half-lives as functions of the $G$ value.}
	\label{fig3}
\end{figure}

In the preliminary calculation of half-lives,
we used $G=20$ and $22$ for small and large $N_{\rm par}$ nuclei as in Ref.~\cite{Choi:2023ztx}, and obtained that the half-lives consistently overestimate the data.
There are many inputs that contribute to the half-lives such as
$Q_\alpha$, $P_\alpha$, $V_l$, $V_C$, $V_N$ and $G$.
$G$ is most flexible among them, so we calculate the root-mean-square deviation defined by 
\begin{eqnarray}
\sigma = \sqrt{\frac{1}{N} \sum^N_{i=1} \left(
\log_{10} T^{\rm cal}_{1/2, i} 
- \log_{10} T^{\rm exp}_{1/2, i} \right)^2}
\label{eq:st}
\end{eqnarray}
for the $G$ values (20, 22), (22, 24), (24, 26) and (26, 28) to determine optimal values of $G$.
The results are shown in Fig. \ref{fig3}.
As the $G$ value increases, $\sigma$ value decreases
and also the dependence on the model becomes weak.
Average value of $\sigma$ with the four models and $G=(24, 26)$ is slightly smaller than that of $G=(26, 28)$,
so in the forthcoming calculations, we fix $G$ values to $24$ for $82 <N_{\rm par} \leq 126$ and $26$ for $N_{\rm par} > 126$.

\begin{figure}
	\centering
	\includegraphics[width=0.45\textwidth]{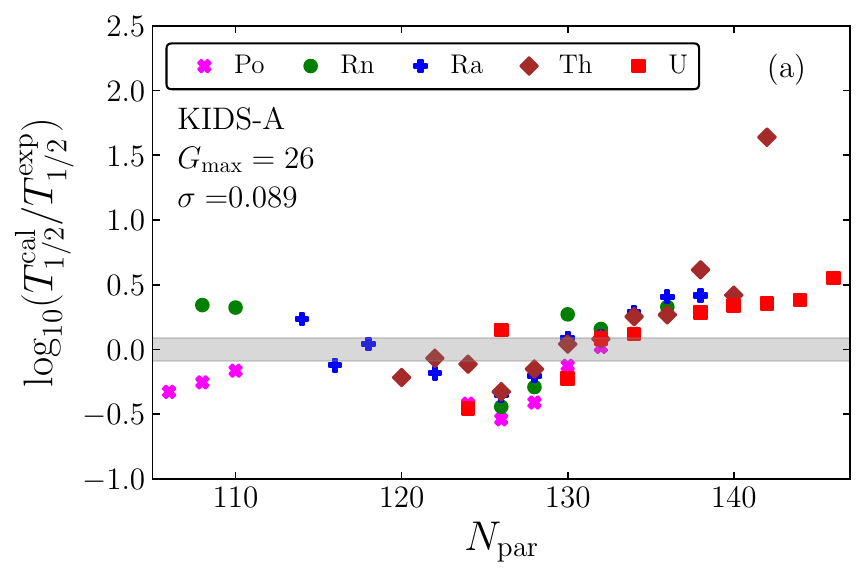}
	\includegraphics[width=0.45\textwidth]{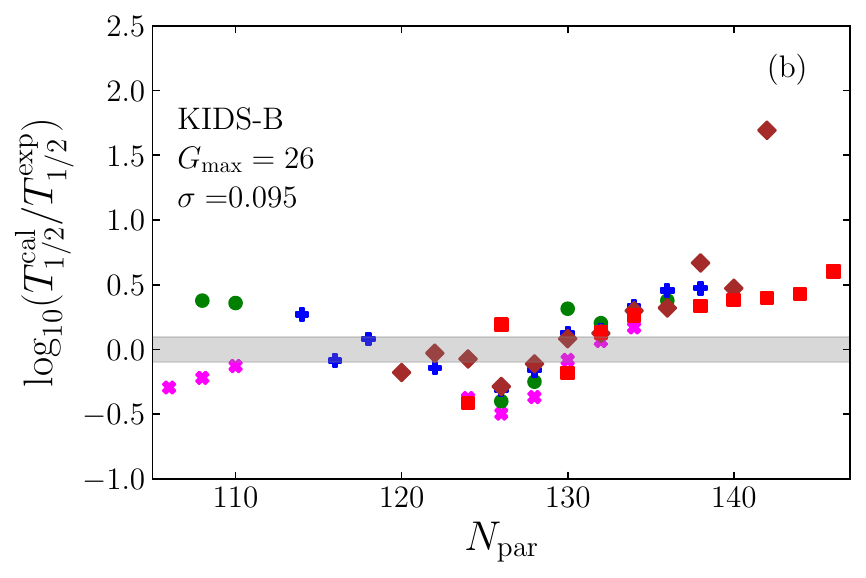}
	\includegraphics[width=0.45\textwidth]{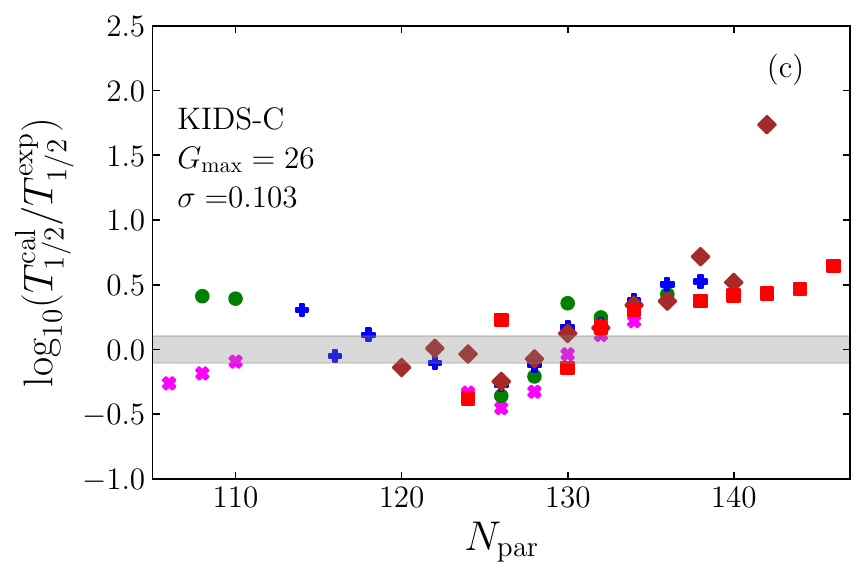}
	\includegraphics[width=0.45\textwidth]{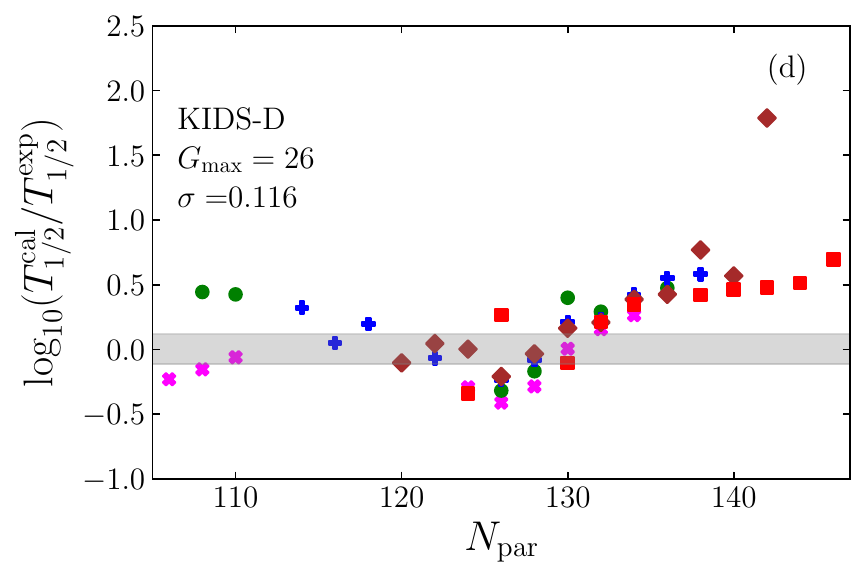}
	\caption{Ratio of half-lives obtained from theory to the experimental values in the logarithmic scale as functions of the neutron number of the parent nuclei.}
	\label{fig4}
\end{figure}

Figure \ref{fig4} illustrates the calculated half-lives divided by the experimental
values in the logarithmic scale as functions of the neutron number of parent nuclei
$N_{\rm par}$ for the four KIDS models.
Almost all the results fall into the range [-0.5, 0.5] in $\log_{10}$ scale.
It corresponds to the factors between 1/3 and 3, so the experimental data
are reproduced reasonably well.
Only $^{232}$Th exhibits overestimation with a factor larger than 10,
but a similar behavior is also obtained with the PC-PK1 model \cite{Choi:2023ztx}.

It is interesting to note that four nuclei except for Uranium show a local
minimum at the magic number $N_{\rm par}=126$.
Below $N_{\rm par}=126$, it is hard to see specific tendency in the behavior 
of the result: As $N_{\rm par}$ increases in $N_{\rm par} \leq 126$, half-lives
of Po isotope increase, but those of Ra isotope decrease and those of
Th show mixed behavior.
On the other hand, for $N_{\rm par} \geq 126$, half-lives increase monotonically
as $N_{\rm par}$ increases for all the nuclides.
Comparing the models, four panels look similar so the dependence on the model
is not seen easily at a glance.
However, comparing the results within or close to the gray shade, one can find
that the half-lives increase systematically from KIDS-A to KIDS-D.
A direct comparison between the models will be considered after we see the 
next result.

\begin{figure}
	\centering
	\includegraphics[width=0.45\textwidth]{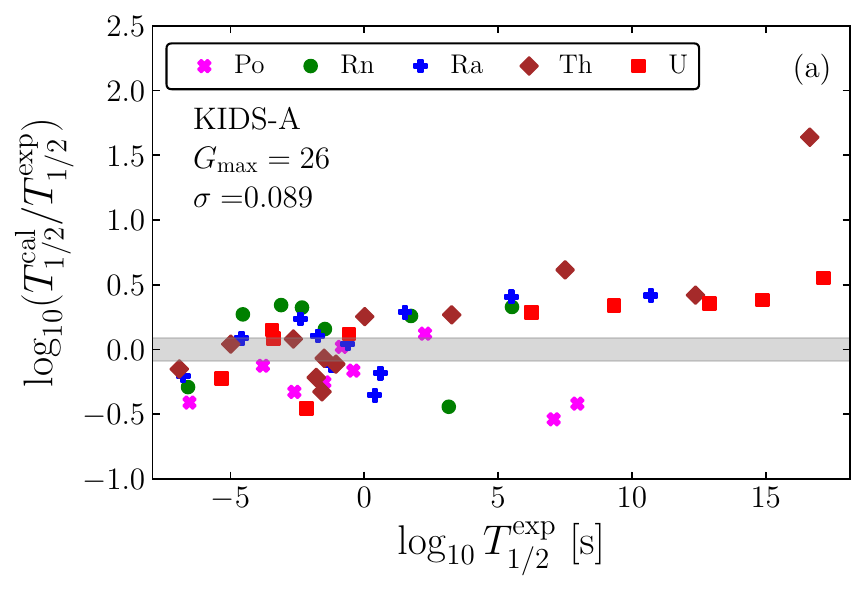}
	\includegraphics[width=0.45\textwidth]{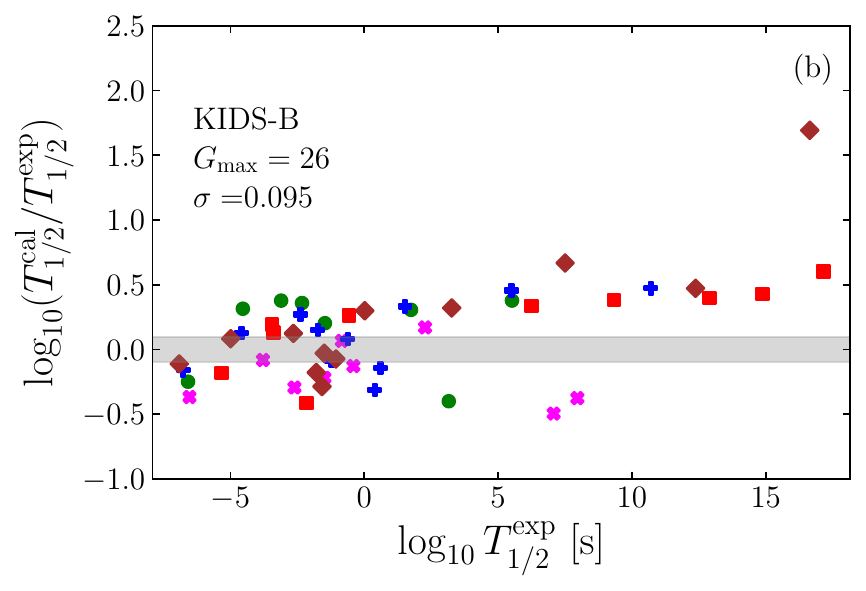}
	\includegraphics[width=0.45\textwidth]{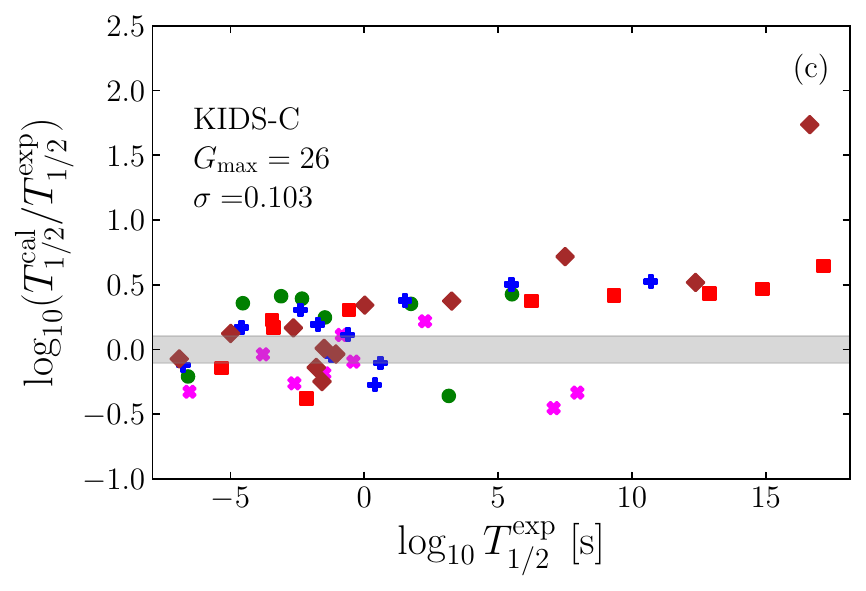}
	\includegraphics[width=0.45\textwidth]{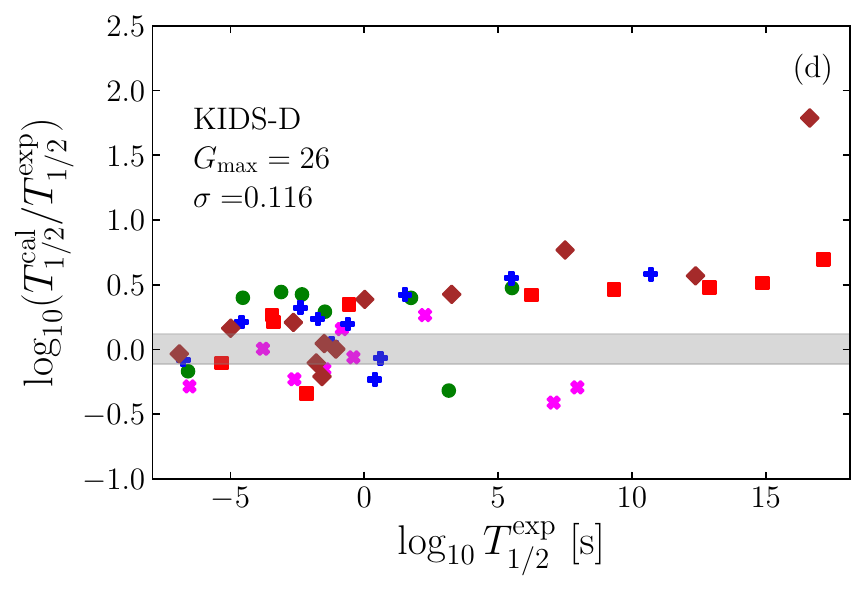}
	\caption{Ratio of half-lives obtained from theory to the experimental values in the logarithmic scale as functions of the experimental half-lives in log scale.}
	\label{fig5}
\end{figure}

In Fig. \ref{fig5}, we analyze the theoretical results as functions of 
experimentally measured half-lives.
Half-lives are in the unit of second, so $\log_{10} T_{1/2} =5$ is close to one day.
For $T_{1/2} < $ 1 day, results are distributed randomly between $-0.5$ and $0.5$, 
and there is no noticeable trend or dependence on $T^{\rm exp}_{1/2}$.
Majority of Po isotopes are below zero and Rn isotopes above zero,
but other isotopes are located uniformly below and above zero.
However, for $T_{1/2}>$ 1 day, except for the Po isotopes,
theoretical results are consistently bigger than the experimental data.
This gives the root-mean-square deviation $\sigma$ of the long-lived isotopes
larger than that of the short-lived ones.
Half-lives become large when the decay width $\Gamma$ in Eq. (\ref{eq:Gamma}) is small.
Small changes in $\Gamma$ caused by $Q_\alpha$ and $V(r, \beta)$ can
lead to large uncertainty in $T_{1/2}$, so relatively large $\sigma$
for long-lived nuclei could be a reasonable result.
However, it is notable that the results in the $T_{1/2}>$ 1 day region
are flat, so the theory prediction is substantially independent of 
$T^{\rm exp}_{1/2}$.
In this plot again, model dependence is not seen clearly,
but it turns out to be transparent and systematic.

\begin{figure}
	\centering
    \includegraphics[width=0.45\textwidth]{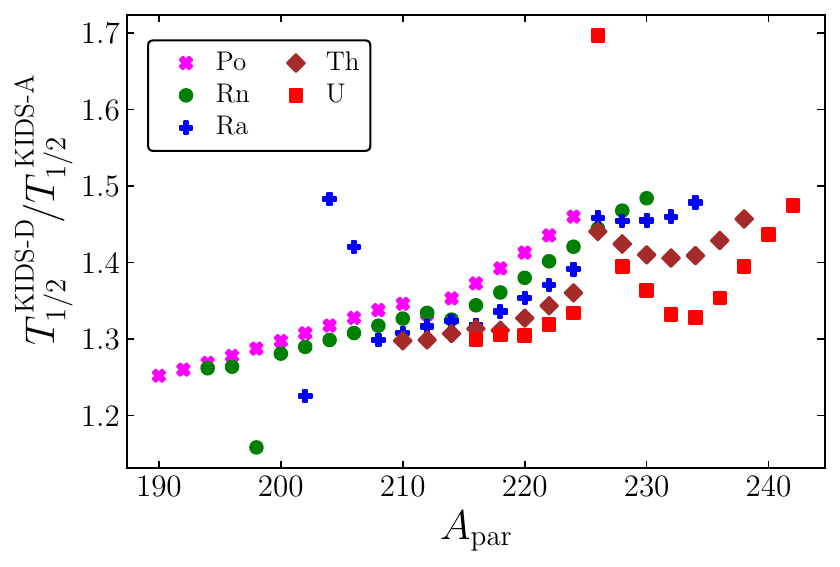}
	\caption{Half-lives obtained from the KIDS-D model divided by those from the 
 KIDS-A model as functions of the mass number of the parent nuclei.}
	\label{fig6}
\end{figure}

In Figs. \ref{fig4} and \ref{fig5}, it is not difficult to see that for a given nucleus,
the half-lives are smallest for the KIDS-A model, and they increase in
the order of KIDS-B, KIDS-C, KIDS-D, and thus the half-lives are longest 
in the KIDS-D model.
Important thing is whether the difference is sizable, and there is a
systematic pattern depending on the symmetry energy.

Figure \ref{fig6} depicts the ratio of half-lives with the KIDS-D model
to those with the KIDS-A model as functions of $A_{\rm par}$. 
Except for $^{198}$Rn and $^{226}$U, the ratios reside in the range 1.2--1.5.
For a fixed $A_{\rm par}$, ratios increase with smaller $Z$ number nuclei,
and for a given nuclear species (fixed $Z$ number), the ratio increases 
monotonically up to $A_{\rm par}=224$.
Combination of these two behaviors gives a result that the ratio increases as 
the neutron number increases, which establishes a straightforward correlation
between the symmetry energy and the alpha-decay half-lives: softer symmetry 
energy (small $J$ and $L$) makes the half-lives longer in more neutron-rich nuclei.
However, the monotonic relationship between the ratio and $A_{\rm par}$ is 
broken for $A_{\rm par} > 224$.
Interestingly local minima are located at $A_{\rm par}=$230, 232 and 234 for
Ra, Th and U, at which the neutron number is identically 
$N=142$.
Small ratio implies that the nuclear structure, especially the neutron distribution
is less dependent on the symmetry energy.
It is worthwhile to investigate why $N=142$ is special in the structure
of neutron-rich heavy nuclei.

A question arises at this point: Is the magnitude 1.2--1.5 large enough to
demonstrate the significance of the symmetry energy in the alpha-decay?
It is well known that the neutron skin thickness of $^{208}$Pb ($\Delta R_{np}$)
is a good indicator to test the stiffness of the symmetry energy.
Accurate measurement of $\Delta R_{np}$ is expected to provide stringent
constraints on the slope parameter $L$ of the symmetry energy.
In order to see how much $\Delta R_{np}$ is sensitive to the symmetry energy,
we calculate it with the KIDS models.
We obtain $\Delta R_{np}=$ 0.235 fm for KIDS-A and 0.168 fm for KIDS-D, and the ratio is 
$\Delta R_{np}(\mbox{KIDS-D})/\Delta R_{np}(\mbox{KIDS-A})=0.71$.
Half-life ratios are similar in magnitude to the inverse of the ratio of the neutron skin 
thickness.
It implies that the alpha-decay is sensitive to the symmetry energy, so
in addition to the neutron skin thickness, the alpha-decay offers a laboratory
to constrain the density dependence of the symmetry energy.
Then, why the half-lives of the soft symmetry energy (KIDS-D) are longer than the
stiff symmetry energy (KIDS-A)?

\begin{figure}
	\centering
    \includegraphics[width=0.45\textwidth]{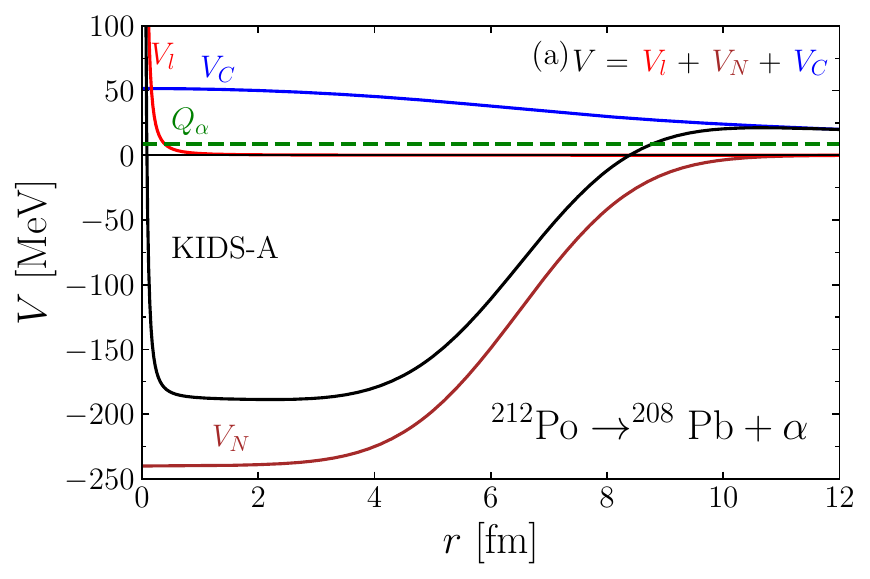}
    \includegraphics[width=0.45\textwidth]{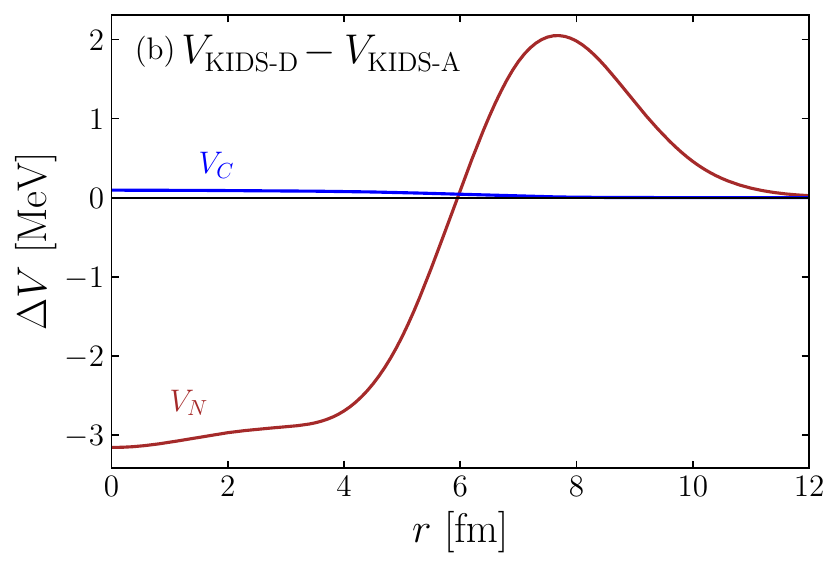}
    \includegraphics[width=0.45\textwidth]{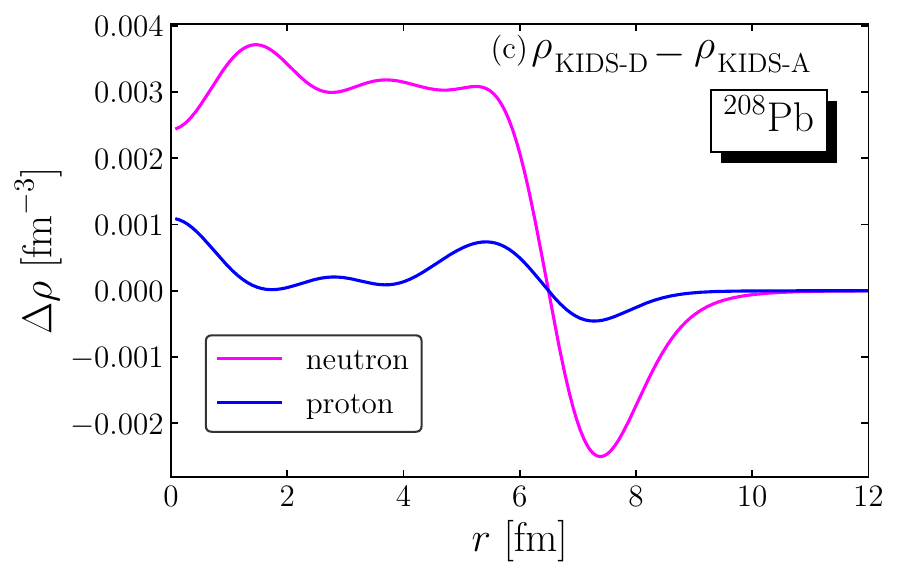}
	\caption{(a) Centrifugal, Coulomb, nuclear potentials and
 their sum in the KIDS-A model. (b) Difference of the Coulomb and the
 nuclear potentials between the KIDS-A and the KIDS-D models. 
 (c) Difference of the neutron and the proton densities between the
 KIDS-A and the KIDS-D models.}
	\label{fig7}
\end{figure}

To understand the origin of the correlation between the symmetry energy
and the alpha-decay half-lives, we compare the potentials.
Fig. \ref{fig7}(a) shows $V_l$, $V_C$, $V_N$ and their sum $V$ of the
KIDS-A model.
Role of $V_l$ is to make a strong repulsion in the core and contributes to $N_f$
by determining $r_1$ in Eq. (\ref{eq:nfapp}).
Tunneling barrier of the alpha-decay is determined by $V_C$ and $V_N$.
The barrier begins around $r\sim8.5$ fm, so the behavior of the matter distribution
in the surface region of a nucleus is critical to the formation and shape of the barrier.

In Fig. \ref{fig7}(b), we plot the differences of the Coulomb $V_C$ and
nuclear $V_N$ potentials between the KIDS-D and the KIDS-A models.
Coulomb potential is identical in the two models, so it is clear that the model dependence
is caused by the nuclear potential.
Since the potential of the KIDS-D model is subtracted by the KIDS-A model, 
negative values in the core ($r \lesssim 6$ fm) means that the nuclear potential of the
KIDS-D model is deeper than the KIDS-A model. In the region $r \gtrsim 6$ fm, 
$V_N$ of the KIDS-D model is larger than the KIDS-A model.
When larger $V_N$ is added to $V_C$, the barrier becomes higher and wider,
so the half-life for KIDS-D becomes longer.

Behavior of $V_C$ and $V_N$ are, as can be seen in Eqs. (\ref{coulV}, \ref{nuclV}),
sensitive to the proton and nucleon distributions in daughter nuclei.
In Fig. \ref{fig7}(c), we compare the proton and the neutron densities
of $^{208}$Pb of the KIDS-D model with those of the KIDS-A model.
We already see in Fig. \ref{fig1} that the proton is insensitive to the symmetry energy.
Explicit calculation of the difference between KIDS-D and KIDS-A confirms the observation
in Fig. \ref{fig1}.
The difference of the neutron density also confirms the result in Fig. \ref{fig1}.
In the core region, the neutron density of the KIDS-D model is enhanced compared to the 
KIDS-A model, so the difference turns out to be positive.
Since the total neutron number is the same in KIDS-D and KIDS-A,
increase of the density in the core region of KIDS-D must be compensated somewhere,
and the result shows that the compensation happens in $6 \lesssim r \lesssim 10$ fm.
Formation of the barrier happens in this region, so the correlation between the symmetry
energy and the alpha-decay half-lives is rooted in the effect of the symmetry energy
to the density distribution in nuclei.

\section{Summary and conclusions} 
\label{summary_and_conclusions}

In this study, we calculated the alpha-decay half-lives of even-even isotopes
with atomic numbers $84 \le Z \le 92$ in the semiclassical WKB approximation framework.
We input the density distribution described by the KIDS models for the density distribution required to calculate the alpha-decay half-lives. 
We adopt cluster-formation model to determine the preformation factor. 
The separation energies for the one proton, one neutron, and $\alpha$ particle are quoted from AME2020 to calculate the preformation factor. 

Before calculating the alpha-decay half-lives, we compared the binding energy and quadrupole deformation calculated from KIDS models with experimental data in order to verify the accuracy of the model. 
Next, we compared the calculated alpha-decay half-lives for different global quantum number $G$ to experimental data. 
If the global quantum number $G$, which is constrained to be only even numbers, is set 20 and 22 as in previous studies~\cite{Choi:2023ztx,Ni:2010zza,Wan:2021wny,Wang:2022axn}, the calculated alpha-decay half-lives in this study are mainly larger than the experimental results.
To reproduce the experimental data of the half-lives more accurately, we set $G$ values as 24 for $82 < N_{\rm par} \leq 126$ and 26 for $N_{\rm par} > 126$.

We then compared the calculated alpha-decay half-lives with experimental data and discussed the results. 
Except for U isotopes, the calculated alpha-decay half-lives at $N_{\rm par} = 126$ are estimated to be smallest compared to the experimental data.
If the measured half-lives are less than about $10^5$ seconds, there is no specific features or patterns. 
However, when the measured half-lives are greater than about $10^5$ seconds, except for the Po isotopes, our calculations of alpha-decay are overestimated as the measured half-lives increase. 
This might be induced by the underestimated decay width $\Gamma$, which is inversely proportional to $T_{1/2}$.

It is noticeable that model predictions become larger if the KIDS model changes from A to D.
For $A_{\rm par} \leq 224$, the ratios of the half-life of the KIDS-D model divided by that of the KIDS-A model generally tends to increase as the number of the neutron increases as in Fig.~\ref{fig6}.
On the other hand, for $A_{\rm par} > 224$,
for  Ra, Th, and U nuclei, the differences between KIDS-D and KIDS-A models become local minima at $N_{\rm par} = 142$.
This implies that the neutron distribution in $N_{\rm par}=142$ is less dependent on the symmetry energy. 
It is necessary to further study its role in $N_{\rm par}=142$ nuclei in heavy nuclei.
We next discussed that the differences of half-life up to 1.5 in ratio between the KIDS-D and the KIDS-A models exhibit strong correlation between the alpha-decay half-lives and the symmetry energy by comparing the neutron skin thickness related to the stiffness of the symmetry energy.

Lastly, we discussed why the half-life varies depending on the KIDS model by 
comparing the nuclear potentials and density distributions obtained by 
the KIDS-D and the KIDS-A models. 
Neutron distribution described by the KIDS-D model is more concentrated in the center of the nucleus than the KIDS-A model. 
This means that the point where the potential barrier and $Q_\alpha$ value meet is closer to the center of the parent nucleus so the width of tunneling barrier becomes wider in the KIDS-D model. 
As a consequence, the alpha-decay half-life in the KIDS-D model becomes greater than the KIDS-A model.


\section*{ACKNOWLEDGMENTS}
This was supported by National Research Foundation funded by Korea Government 
(Grant Nos. 2018R1A5A1025563, 2023R1A2C1003177, and 2023R1A2C1005398).
This work was supported by the National Supercomputing Center with supercomputing resources including technical support (KSC-2023-CRE-0170, KSC-2022-CRE-0373, and KSC-2023-CRE-0521).

\bibliographystyle{aip}

\pagebreak

\appendix
\onecolumngrid

\section{Tables of the numerical results}

\begin{table*}[h] 
\centering
\caption{Alpha-decay half-lives of the Po isotopes calculated with the KIDS-A, KIDS-B, KIDS-C and KIDS-D models. 
Alpha particle energy $Q_\alpha$ and the preformation factor $P_\alpha$ are calculated from the binding energies from AME2020. \label{tab2} }
\begin{tabular}{c|c|c|c|c|c|c|c|c|c|c|c}\hline
Parent & \multirow{2}{*}{$Q_\alpha$} & \multicolumn{4}{c|}{$\beta_{\rm 2, par}$} & \multirow{2}{*}{$P_\alpha$}  & \multicolumn{5}{c}{$\log T_{1/2}$} \\ \cline{3-6} \cline{8-12}
nucleus  &       & KIDS-A & KIDS-B & KIDS-C & KIDS-D &         & Exp. & KIDS-A & KIDS-B & KIDS-C & KIDS-D \\ \hline
$^{190}$Po &  7.693 & -0.080 & -0.080 & -0.079 & -0.079 &  0.262 & -2.611 & -2.939 & -2.905 & -2.873 & -2.842  \\
$^{192}$Po &  7.320 & -0.076 & -0.077 & -0.077 & -0.077 &  0.257 & -1.492 & -1.746 & -1.712 & -1.678 & -1.646  \\
$^{194}$Po &  6.987 & 0.000 & -0.070 & -0.071 & -0.072 &  0.235 & -0.407 & -0.571 & -0.535 & -0.501 & -0.468  \\
$^{196}$Po &  6.658 & 0.000 & 0.000 & 0.000 & 0.000 &  0.216 &  &  0.679 &  0.716 &  0.751 &  0.785  \\
$^{198}$Po &  6.310 & 0.000 & 0.002 & 0.000 & 0.000 &  0.203 &  &  2.104 &  2.142 &  2.178 &  2.214  \\
$^{200}$Po &  5.982 & 0.000 & 0.000 & 0.000 & 0.000 &  0.184 &  &  3.580 &  3.619 &  3.656 &  3.693  \\
$^{202}$Po &  5.701 & 0.000 & 0.000 & 0.000 & 0.000 &  0.177 &  &  4.920 &  4.960 &  4.999 &  5.037  \\
$^{204}$Po &  5.485 & 0.000 & 0.000 & 0.000 & 0.000 &  0.158 &  &  6.052 &  6.093 &  6.133 &  6.172  \\
$^{206}$Po &  5.327 & 0.000 & 0.000 & 0.000 & 0.000 &  0.145 &  &  6.915 &  6.957 &  6.998 &  7.038  \\
$^{208}$Po &  5.216 & 0.000 & 0.000 & 0.000 & 0.000 &  0.135 & 7.961 &  7.542 &  7.585 &  7.626 &  7.668  \\
$^{210}$Po &  5.408 & 0.000 & 0.000 & 0.000 & 0.000 &  0.105 & 7.078 &  6.537 &  6.581 &  6.624 &  6.666  \\
$^{212}$Po &  8.954 & 0.000 & 0.000 & 0.000 & 0.000 &  0.221 & -6.531 & -6.942 & -6.900 & -6.858 & -6.817  \\
$^{214}$Po &  7.834 & 0.000 & 0.000 & 0.000 & 0.000 &  0.213 & -3.787 & -3.913 & -3.868 & -3.825 & -3.782  \\
$^{216}$Po &  6.906 & 0.000 & 0.000 & 0.000 & 0.000 &  0.205 & -0.842 & -0.822 & -0.776 & -0.730 & -0.685  \\
$^{218}$Po &  6.115 & 0.000 & 0.000 & 0.000 & 0.000 &  0.196 & 2.269 &  2.390 &  2.439 &  2.486 &  2.534  \\
$^{220}$Po &  5.329 & 0.000 & 0.000 & 0.000 & 0.000 &  0.215 &  &  6.241 &  6.292 &  6.341 &  6.391  \\
$^{222}$Po &  4.432 & 0.000 & 0.000 & 0.000 & 0.000 &  0.200 &  & 11.998 & 12.051 & 12.103 & 12.155  \\
$^{224}$Po &  3.355 & 0.000 & 0.000 & 0.000 & 0.000 &  0.176 &  & 22.368 & 22.422 & 22.478 & 22.532  \\
\hline
\end{tabular}
\end{table*}

\begin{table*}[h] 
\centering
\caption{Alpha-decay half-lives of the Rn isotopes calculated with the KIDS-A, KIDS-B, KIDS-C and KIDS-D models. 
Alpha particle energy $Q_\alpha$ and the preformation factor $P_\alpha$ are calculated from the binding energies from AME2020. \label{tab3} }
\begin{tabular}{c|c|c|c|c|c|c|c|c|c|c|c}\hline
Parent & \multirow{2}{*}{$Q_\alpha$} & \multicolumn{4}{c|}{$\beta_{\rm 2, par}$} & \multirow{2}{*}{$P_\alpha$}  & \multicolumn{5}{c}{$\log T_{1/2}$} \\ \cline{3-6} \cline{8-12}
nucleus  &       & KIDS-A & KIDS-B & KIDS-C & KIDS-D &         & Exp. & KIDS-A & KIDS-B & KIDS-C & KIDS-D \\ \hline
$^{194}$Rn &  7.862 & 0.146 & 0.146 & 0.149 & 0.148 &  0.261 & -3.108 & -2.766 & -2.731 & -2.697 & -2.665  \\
$^{196}$Rn &  7.617 & -0.120 & 0.129 & 0.132 & 0.131 &  0.256 & -2.328 & -2.005 & -1.970 & -1.936 & -1.903  \\
$^{198}$Rn &  7.349 & -0.109 & -0.109 & -0.110 & 0.114 &  0.239 &  & -1.064 & -1.067 & -1.033 & -1.000  \\
$^{200}$Rn &  7.043 & -0.095 & -0.095 & -0.097 & -0.097 &  0.221 &  &  0.031 &  0.068 &  0.093 &  0.139  \\
$^{202}$Rn &  6.774 & -0.076 & -0.077 & -0.080 & -0.080 &  0.209 &  &  1.050 &  1.088 &  1.125 &  1.161  \\
$^{204}$Rn &  6.547 & 0.000 & 0.000 & 0.000 & 0.000 &  0.191 &  &  1.973 &  2.012 &  2.049 &  2.086  \\
$^{206}$Rn &  6.384 & 0.000 & 0.000 & 0.000 & 0.000 &  0.180 &  &  2.653 &  2.693 &  2.732 &  2.770  \\
$^{208}$Rn &  6.261 & 0.000 & 0.000 & 0.000 & 0.000 &  0.163 &  &  3.202 &  3.243 &  3.320 &  3.322  \\
$^{210}$Rn &  6.159 & 0.000 & 0.000 & 0.000 & 0.000 &  0.151 &  &  3.656 &  3.698 &  3.739 &  3.779  \\
$^{212}$Rn &  6.385 & 0.000 & 0.000 & 0.000 & 0.000 &  0.121 & 3.157 &  2.713 &  2.756 &  2.797 &  2.838  \\
$^{214}$Rn &  9.208 & 0.000 & 0.000 & 0.000 & 0.000 &  0.228 & -6.587 & -6.878 & -6.836 & -6.796 & -6.756  \\
$^{216}$Rn &  8.198 & 0.000 & 0.000 & 0.000 & 0.000 &  0.237 & -4.538 & -4.267 & -4.223 & -4.181 & -4.139  \\
$^{218}$Rn &  7.262 & 0.001 & 0.000 & 0.000 & 0.000 &  0.234 & -1.472 & -1.314 & -1.269 & -1.225 & -1.181  \\
$^{220}$Rn &  6.405 & 0.000 & 0.000 & 0.000 & 0.000 &  0.221 & 1.745 &  2.003 &  2.050 &  2.097 &  2.143  \\
$^{222}$Rn &  5.590 & 0.115 & 0.113 & 0.112 & 0.111 &  0.222 & 5.519 &  5.846 &  5.896 &  5.945 &  5.993  \\
$^{224}$Rn &  4.757 & 0.131 & 0.130 & 0.130 & 0.130 &  0.214 &  & 10.836 & 10.887 & 10.938 & 10.989  \\
$^{226}$Rn &  3.836 & 0.143 & 0.144 & 0.144 & 0.145 &  0.204 &  & 18.217 & 18.271 & 18.324 & 18.377  \\
$^{228}$Rn &  2.908 & 0.155 & 0.155 & 0.157 & 0.157 &  0.201 &  & 31.146 & 31.202 & 31.257 & 31.313  \\
$^{230}$Rn &  2.196 & 0.166 & 0.167 & 0.168 & 0.169 &  0.154 &  & 44.287 & 44.347 & 44.405 & 44.458  \\
\hline
\end{tabular}
\end{table*}

\begin{table*}[h] 
\centering
\caption{Alpha-decay half-lives of the Ra isotopes calculated with the KIDS-A, KIDS-B, KIDS-C and KIDS-D models. 
Alpha particle energy $Q_\alpha$ and the preformation factor $P_\alpha$ are calculated from the binding energies from AME2020. \label{tab4} }
\begin{tabular}{c|c|c|c|c|c|c|c|c|c|c|c}\hline
Parent & \multirow{2}{*}{$Q_\alpha$} & \multicolumn{4}{c|}{$\beta_{\rm 2, par}$} & \multirow{2}{*}{$P_\alpha$}  & \multicolumn{5}{c}{$\log T_{1/2}$} \\ \cline{3-6} \cline{8-12}
nucleus  &       & KIDS-A & KIDS-B & KIDS-C & KIDS-D &         & Exp. & KIDS-A & KIDS-B & KIDS-C & KIDS-D \\ \hline
$^{202}$Ra &  7.880 & 0.134 & 0.134 & 0.134 & 0.133 &  0.248 & -2.387 & -2.153 & -2.117 & -2.083 & -2.065  \\
$^{204}$Ra &  7.637 & 0.111 & 0.111 & 0.112 & 0.111 &  0.231 & -1.222 & -1.345 & -1.308 & -1.274 & -1.174  \\
$^{206}$Ra &  7.415 & 0.081 & 0.081 & -0.082 & -0.082 &  0.219 & -0.62 & -0.577 & -0.540 & -0.506 & -0.424  \\
$^{208}$Ra &  7.273 & 0.000 & 0.000 & 0.000 & 0.000 &  0.196 &  & -0.015 &  0.024 &  0.062 &  0.099  \\
$^{210}$Ra &  7.151 & 0.000 & 0.000 & 0.000 & 0.000 &  0.187 & 0.602 &  0.417 &  0.458 &  0.496 &  0.534  \\
$^{212}$Ra &  7.032 & 0.000 & 0.000 & 0.000 & 0.000 &  0.171 &  &  0.870 &  0.911 &  0.950 &  0.989  \\
$^{214}$Ra &  7.273 & 0.000 & 0.000 & 0.000 & 0.000 &  0.138 & 0.387 &  0.032 &  0.074 &  0.113 &  0.154  \\
$^{216}$Ra &  9.526 & 0.000 & 0.000 & 0.000 & 0.000 &  0.239 & -6.764 & -6.967 & -6.926 & -6.887 & -6.847  \\
$^{218}$Ra &  8.540 & 0.000 & 0.000 & 0.000 & 0.000 &  0.243 & -4.587 & -4.501 & -4.459 & -4.417 & -4.376  \\
$^{220}$Ra &  7.594 & 0.000 & 0.000 & 0.000 & 0.000 &  0.239 & -1.742 & -1.638 & -1.593 & -1.550 & -1.507  \\
$^{222}$Ra &  6.678 & 0.121 & 0.119 & 0.116 & 0.114 &  0.199 & 1.526 &  1.812 &  1.858 &  1.904 &  1.949  \\
$^{224}$Ra &  5.789 & 0.142 & 0.141 & 0.141 & 0.140 &  0.184 & 5.497 &  5.903 &  5.952 &  5.999 &  6.046  \\
$^{226}$Ra &  4.871 & 0.158 & 0.158 & 0.159 & 0.158 &  0.182 & 10.703 & 11.119 & 11.175 & 11.227 & 11.283  \\
$^{228}$Ra &  4.070 & 0.171 & 0.172 & 0.174 & 0.175 &  0.184 &  & 17.209 & 17.265 & 17.317 & 17.372  \\
$^{230}$Ra &  3.344 & 0.183 & 0.184 & 0.187 & 0.187 &  0.167 &  & 25.602 & 25.659 & 25.710 & 25.765  \\
$^{232}$Ra &  2.829 & 0.194 & 0.195 & 0.197 & 0.198 &  0.152 &  & 34.190 & 34.247 & 34.299 & 34.354  \\
$^{234}$Ra &  2.336 & 0.203 & 0.204 & 0.206 & 0.207 &  0.132 &  & 43.240 & 43.296 & 43.353 & 43.410  \\
\hline
\end{tabular}
\end{table*}

\begin{table*}[h] 
\centering
\caption{Alpha-decay half-lives of the Th isotopes calculated with the KIDS-A, KIDS-B, KIDS-C and KIDS-D models. 
Alpha particle energy $Q_\alpha$ and the preformation factor $P_\alpha$ are calculated from the binding energies from AME2020. \label{tab5} }
\begin{tabular}{c|c|c|c|c|c|c|c|c|c|c|c}\hline
Parent & \multirow{2}{*}{$Q_\alpha$} & \multicolumn{4}{c|}{$\beta_{\rm 2, par}$} & \multirow{2}{*}{$P_\alpha$}  & \multicolumn{5}{c}{$\log T_{1/2}$} \\ \cline{3-6} \cline{8-12}
nucleus  &       & KIDS-A & KIDS-B & KIDS-C & KIDS-D &         & Exp. & KIDS-A & KIDS-B & KIDS-C & KIDS-D \\ \hline
$^{210}$Th &  8.069 & 0.000 & 0.068 & -0.071 & -0.071 &  0.234 & -1.796 & -2.013 & -1.975 & -1.936 & -1.900  \\
$^{212}$Th &  7.958 & 0.000 & 0.000 & 0.000 & 0.000 &  0.196 & -1.499 & -1.567 & -1.528 & -1.491 & -1.454  \\
$^{214}$Th &  7.827 & 0.000 & 0.000 & 0.000 & 0.000 &  0.195 & -1.06 & -1.175 & -1.135 & -1.096 & -1.058  \\
$^{216}$Th &  8.072 & 0.000 & 0.000 & 0.000 & 0.000 &  0.159 & -1.58 & -1.908 & -1.867 & -1.828 & -1.790  \\
$^{218}$Th &  9.849 & 0.000 & 0.000 & 0.000 & 0.000 &  0.251 & -6.914 & -7.066 & -7.025 & -6.987 & -6.948  \\
$^{220}$Th &  8.973 & 0.000 & 0.000 & 0.000 & 0.000 &  0.243 & -4.991 & -4.951 & -4.909 & -4.868 & -4.828  \\
$^{222}$Th &  8.133 & 0.100 & 0.000 & 0.000 & 0.000 &  0.232 & -2.65 & -2.570 & -2.526 & -2.484 & -2.441  \\
$^{224}$Th &  7.299 & 0.138 & 0.136 & 0.135 & 0.134 &  0.198 & 0.017 &  0.270 &  0.315 &  0.359 &  0.403  \\
$^{226}$Th &  6.453 & 0.160 & 0.160 & 0.162 & 0.162 &  0.182 & 3.265 &  3.532 &  3.585 &  3.638 &  3.691  \\
$^{228}$Th &  5.520 & 0.178 & 0.180 & 0.183 & 0.184 &  0.183 & 7.504 &  8.119 &  8.172 &  8.221 &  8.272  \\
$^{230}$Th &  4.770 & 0.195 & 0.197 & 0.201 & 0.202 &  0.183 & 12.376 & 12.796 & 12.848 & 12.894 & 12.945  \\
$^{232}$Th &  4.082 & 0.209 & 0.211 & 0.215 & 0.216 &  0.160 & 16.645 & 18.285 & 18.338 & 18.381 & 18.433  \\
$^{234}$Th &  3.672 & 0.220 & 0.222 & 0.225 & 0.225 &  0.151 &  & 22.604 & 22.658 & 22.701 & 22.753  \\
$^{236}$Th &  3.333 & 0.228 & 0.229 & 0.231 & 0.231 &  0.140 &  & 27.302 & 27.356 & 27.404 & 27.457  \\
$^{238}$Th &  3.169 & 0.232 & 0.233 & 0.234 & 0.234 &  0.135 &  & 29.875 & 29.931 & 29.984 & 30.039  \\
\hline
\end{tabular}
\end{table*}

\begin{table*}[h] 
\centering
\caption{Alpha-decay half-lives of the U isotopes calculated with the KIDS-A, KIDS-B, KIDS-C and KIDS-D models. 
Alpha particle energy $Q_\alpha$ and the preformation factor $P_\alpha$ are calculated from the binding energies from AME2020. \label{tab6} }
\begin{tabular}{c|c|c|c|c|c|c|c|c|c|c|c}\hline
Parent & \multirow{2}{*}{$Q_\alpha$} & \multicolumn{4}{c|}{$\beta_{\rm 2, par}$} & \multirow{2}{*}{$P_\alpha$}  & \multicolumn{5}{c}{$\log T_{1/2}$} \\ \cline{3-6} \cline{8-12}
nucleus  &       & KIDS-A & KIDS-B & KIDS-C & KIDS-D &         & Exp. & KIDS-A & KIDS-B & KIDS-C & KIDS-D \\ \hline
$^{216}$U &  8.531 & 0.000 & 0.000 & 0.000 & 0.000 &  0.206 & -2.161 & -2.616 & -2.577 & -2.539 & -2.502  \\
$^{218}$U &  8.775 & 0.000 & 0.000 & 0.000 & 0.000 &  0.183 & -3.451 & -3.301 & -3.261 & -3.223 & -3.185  \\
$^{220}$U & 10.290 & 0.000 & 0.000 & 0.000 & 0.000 &  0.261 &  & -7.412 & -7.372 & -7.334 & -7.296  \\
$^{222}$U &  9.481 & 0.000 & 0.000 & 0.000 & 0.000 &  0.245 & -5.328 & -5.552 & -5.511 & -5.472 & -5.432  \\
$^{224}$U &  8.628 & 0.112 & 0.000 & 0.105 & 0.000 &  0.247 & -3.402 & -3.317 & -3.274 & -3.233 & -3.192  \\
$^{226}$U &  7.701 & 0.153 & 0.152 & 0.154 & 0.154 &  0.207 & -0.57 & -0.453 & -0.309 & -0.266 & -0.223  \\
$^{228}$U &  6.800 & 0.179 & 0.181 & 0.185 & 0.186 &  0.191 &  &  2.888 &  2.938 &  2.984 &  3.032  \\
$^{230}$U &  5.992 & 0.203 & 0.205 & 0.210 & 0.210 &  0.187 & 6.243 &  6.529 &  6.577 &  6.617 &  6.664  \\
$^{232}$U &  5.414 & 0.222 & 0.224 & 0.227 & 0.227 &  0.169 & 9.337 &  9.674 &  9.719 &  9.753 &  9.799  \\
$^{234}$U &  4.858 & 0.236 & 0.237 & 0.239 & 0.239 &  0.150 & 12.889 & 13.244 & 13.287 & 13.320 & 13.367  \\
$^{236}$U &  4.573 & 0.245 & 0.245 & 0.246 & 0.246 &  0.153 & 14.869 & 15.251 & 15.297 & 15.334 & 15.382  \\
$^{238}$U &  4.270 & 0.250 & 0.250 & 0.250 & 0.250 &  0.137 & 17.149 & 17.700 & 17.750 & 17.792 & 17.845  \\
$^{240}$U &  4.035 & 0.252 & 0.252 & 0.252 & 0.251 &  0.146 &  & 19.729 & 19.782 & 19.831 & 19.886  \\
$^{242}$U &  3.670 & 0.252 & 0.252 & 0.251 & 0.251 &  0.163 &  & 23.867 & 23.924 & 23.978 & 24.036  \\
\hline
\end{tabular}
\end{table*}

\end{document}